# Sensing electric fields using single diamond spins

F. Dolde [1], H. Fedder [1], M.W. Doherty[2], T. Nöbauer [3], F. Rempp [1], G. Balasubramanian [1], T.Wolf [1], F. Reinhard [1], L.C.L. Hollenberg[2], F. Jelezko [1] and J. Wrachtrup [1]

[1] 3rd Institute of Physicsand Research Center SCOPE, University Stuttgart, Pfaffenwaldring 57, D-70550 Stuttgart, Germany

[2]Centre for Quantum Computer Technology, School of Physics, University of Melbourne, Victoria 3010, Australia

[3] Atominstitut TU Wien, Stadionalleee 2, 1020 Wien, Austria

**The ability to sensitively detect charges under ambient conditions would be a fascinating new tool benefitting a wide range of researchers across disciplines. However, most current techniques are limited to low-temperature methods like single-electron transistors (SET)[1,2], single–electron electrostatic force microscopy[3] and scanning tunnelling microscopy [4]. Here we open up a new quantum metrology technique demonstrating precision electric field measurement using a single nitrogen-vacancy defect centre(NV) spin in diamond. An AC electric field sensitivity reaching ~ 140V/cm/√Hz has been achieved. This corresponds to the electric field produced by a single elementary charge located at a distance of ~ 150 nm from our spin sensor with averaging for one second. By careful analysis of the electronic structure of the defect centre, we show how an applied magnetic field influences the electric field sensing properties. By this we demonstrate that diamond defect centre spins can be switched between electric and magnetic field sensing modes and identify suitable parameter ranges for both detector schemes. By combining magnetic and electric field sensitivity, nanoscale detection and ambient operation our study opens up new frontiers in imaging and sensing applications ranging from material science to bioimaging.**

Sensitive imaging or detection of charges is an outstanding task in a variety of applications. The development of e.g. single electron transistors (SET)[5] has pushed charge sensing to an unprecedented sensitivity of $10^{-6}$electron charge, being used in low temperature detection and scanning applications[2] or as sensors in quantum devices[6]. Inspired by the development of tunnelling microscopy a variety of scanning probes have been devised to measure surface electrical properties such as capacitance microscopy(SCM) [7], scanning Kelvin probe (Kelvin Probe) [8] and electric field-sensitive atomic force microscopy(EFM) [9]. Indeed the last method has shown the remarkable ability to detect the presence of individual charges [3].

We report on a fundamentally new method which uses the spin of single defect centres in diamond to sense electric field dependent shifts in energy levels. Sensitive electric field detection is based on the remarkable properties of the NV centre [10]. Most notably these are that the fluorescence of single defects can be detected making it an atom sized local probe [11], the outstandingly long spin dephasing times [12] as well as the controlled positioning of single centres [13]. These properties have lead to a variety of applications of the centre ranging from quantum science [14], precision magnetic field sensing [15,16,17,18,19] to bio labelling. [20,21,22]

The NV centre is a well-studied defect in diamond consisting of a substitutional nitrogen atom adjacent to a carbon vacancy (see figure 1a). The ground spin triplet state exhibits a zero-field splitting between the $m_s = 0$ and degenerate $m_s = \pm 1$ spin sub-levels of $D_{gs} \approx 2.87$ GHz. The spin state of the triplet can be polarized and readout at room-temperature via optical excitation to the excited triplet state, thereby enabling the implementation of optically detected magnetic resonance (ODMR) techniques using microwave sources tuned to the ground state zero-field splitting.[11]The Stark shift in the excited triplet states of single defect centres have been previously documented at low temperature[23], and the ground state Stark shift of NV ensembles has been demonstrated at low temperature [24].This work demonstrates how improved ODMR and magnetic field alignment techniques provide enhanced sensitivity and the means to determine both the magnitude and three dimensional orientation (up to a four-fold symmetry) of the electric field in the vicinity of single NV centres in their ground state.

In our experiments, the electric field was generated by the application of a controlled voltage to a gold microstructure fabricated by lithography and electroplating directly on a bulk diamond sample containing NV centres (see figure 1b). A particular NV centre within the gold microstructure was identified and used for all of our measurements. Two Helmholtz coil pairs for the x-y axes and a single coil for the z axis were used to generate and precisely orientate the applied magnetic field. The resonant microwave field was generated by a set of coils fabricated as part of the gold microstructure. To measure electric field induced shifts of spin sub-levels we optically detected the electron spin resonance transition between the $m_s=0$ and $m_s=\pm1$ sub-levels in the triplet ground state of the centre. Upon application of an electric field of ca. 1000 V/cm resonance line shifts of 28.4 kHz were observed. Before proceeding into the detailed description of the ground state Stark effect, we first analyse the spin-Hamiltonian of the ground triplet state in the presence of applied magnetic, strain and electric fields

$$H_{gs} = (D_{gs} + d_{gs}^{\parallel}\Pi_z)[S_z^2 - 1/3 S(S+1)] + \mu_B g_e \vec{S}\cdot\vec{B} - d_{gs}^{\perp}[\Pi_x(S_xS_y + S_yS_x) + \Pi_y(S_x^2 - S_y^2)] \quad (1)$$

Here $d_{gs}^{\parallel} = 0.35 \pm 0.02$ Hz cm/V and $d_{gs}^{\perp} = 17 \pm 2.5$ Hz cm/V [24] are the measured axial and non-axial Stark shift components of the ground triplet state permanent electric dipole moment $\vec{d}_{gs}$, $\mu_B$ is the Bohr magneton, $g_e$ is the electron g-factor, $\vec{B}$ is the applied magnetic field, and $\vec{S}$ is the electron spin operator.It is worth mentioning that $\vec{d}_{gs}$, is rather small as compared to the excited state $\vec{d}_{es}$ because of the antisymmetric combination of molecular $e_x$, $e_y$ type orbitals in the ground state. In addition contributions from the excited state are reduced by the large energy gap $E_{es}$ between excited and ground state to $\vec{d}_{gs} = \lambda_\perp^2/E_{es} \vec{d}_{es}$, where $\lambda_\perp^2$ is the excited state spin orbit coupling.(see [25]and supplementary information).Given that crystal strain can be treated effectively as a local static electric field $\vec{\sigma}$, [23] strain and the applied electric field $\vec{E}$ combine to form the total effective electric field $\vec{\Pi} = \vec{E} + \vec{\sigma}$ at the NV centre. Note that as depicted in figure 2athe coordinate system is defined such that the z axis is parallel to the axial symmetry axis of the NV centre which connects the nitrogen and vacancy sites.

Figure 2b shows the electric field induced shifts as a function of the applied magnetic field in the z direction. Because the electric field induced shifts are small compared to the Zeeman effect the interplay between the magnetic and electric fields in (1) needs to be analysed. In the regime where $D_{gs} \gg \mu_B g_e B$ and $D_{gs} \gg d_{gs}\Pi$, second-order perturbation theory can be applied to obtain the approximate energy solutions of $H_{gs}$. Considering fixed magnetic and strain fields, the change in the magnetic transition frequency $\Delta\omega_\pm$ between the $m_s = 0$ and the $m_s = \pm1$ spin sub-levels caused by the application of an electric field is

$$\hbar\Delta\omega_\pm = d_{gs}^{\parallel}\Pi_z \pm [F(\vec{B},\vec{E},\vec{\sigma}) - F(\vec{B},\vec{0},\vec{\sigma})] \quad (2)$$

where

$$F(\vec{B}, \vec{E}, \vec{\sigma}) = \left[\mu_B^2 g_e^2 B_z^2 + d_{gs}^{\perp 2}\Pi_\perp^2 - \frac{\mu_B^2 g_e^2}{D_{gs}} B_\perp^2 d_{gs}^\perp (\Pi_x \cos 2\phi_B - \Pi_y \sin 2\phi_B) + \frac{\mu_B^4 g_e^4}{4D_{gs}^2} B_\perp^4\right]^{1/2} \quad (3)$$

and $\tan\phi_B = B_y/B_x$. Given that the Zeeman effect is much stronger than the Stark effect in the ground state, the first-order appearance of $\mu_B^2 g_e^2 B_z^2$ in $F$ indicates that for the expected case $d_{gs}\Pi/\mu_B g_e B \ll 1$, careful alignment of the magnetic field in the non-axial plane is required, otherwise the detectable change in the transition frequency caused by an electric field decays to $\hbar\Delta\omega \approx d_{gs}^\parallel \Pi_z$, which provides no information about the electric field in the non-axial plane and, as $d_{gs}^\parallel$ is small, severely reduces the detectable electric field strength. This argument also suggests that the presence of non-axial strain increases the visibility of the non-axial electric field by increasing the ratio $d_{gs}\Pi/\mu_B g_e B$.

The dependence of $F$ on the non-axial magnetic field orientation $\phi_B$ indicates that the orientation of the electric field will be determined up to a four-fold symmetry if $\Delta\omega$ is measured for several magnetic field orientations. Such a sequence of measurements requires precise control of the magnetic field as it is rotated adiabatically in the non-axial plane. Figure 2c contains the theoretical polar plot of $\Delta\omega$ as a function of $\phi_B$ for the special case (blue line) where the effective strain and applied electric fields are in parallel directions or negligible strain exists. Since $\Delta\omega$ depends on both the orientations of the effective strain and applied electric fields, the symmetric 'four-leaf' pattern of the aligned case will distort and rotate if the applied electric field is rotated with respect to the strain field (red dashed line). If the strain field is characterized for a given NV centre, the measurement of the $\Delta\omega$ polar pattern and subsequent analysis using equation (2), allows the applied electric field to be determined.

Indeed the measured polar plot of $\Delta\omega$ as a function of $\phi_B$ (depicted in figure 2d) agrees very well with the theoretical analysis. The small differences between the theory and the measurements may be explained by a small axial magnetic field uncertainty of ±0.003 mT that was present. The rotation and distortion of the polar pattern from the symmetric 'four-leaf' pattern indicates that the applied electric and effective strain fields are not aligned. A least square fit of the polar pattern yielded the field parameters: $B_\perp = 23.6 \pm 1.51$ G, $d_{gs}^\parallel E_z = -4.19 \pm 2.11$ kHz, $d_{gs}^\perp E_\perp = 81.6 \pm 1.74$ kHz, $\phi_E = 32° \pm 1.16°$, and $\phi_\sigma = 22° \pm 8.62°$. Note that the non-axial orientations of the electric $\phi_E$ and strain $\phi_\sigma$ fields are taken with respect to the laboratory reference frame. The observed strong decay of the electric field induced shift $\Delta\omega$ with increasing axial magnetic field strength (depicted in figure 2b) also agrees well with the theoretical model and reinforces the requirement for precise alignment of the magnetic field in the non-axial plane. The measurements of figures 2b and 2d and the underpinning theoretical analysis, clearly demonstrates how the combination of ODMR techniques and precise magnetic field alignment and manipulation can be used to detect the vector electric field.

As in magnetic field sensing [16] electric fields are most sensitively detected when a field induced phase accumulation is used [25]. As an example figure 3c shows the amplitude of a Hahn echo $\propto \cos(\Delta\omega\tau)$ as a function of electric field strength recorded by a two pulse spin echo sequence with a fixed free evolution time τ. The sinusoidal modulation of the measured echo intensity upon the increase of the electric field allows a precise determination of the minimally detected electric field $\delta E_{min}$. $\delta E_{min}$ can be expressed as the ratio of the standard deviation of a Hahn signal measurement $\sigma_{sn}$ and the gradient of the oscillation of the Hahn signal $\delta S$ (as indicated in figure 3c),

$$\delta E_{min} = \sigma_{sn}/\delta S \qquad (4)$$

In our measurement, $\sigma_{sn}$ depends on the photon shot noise, enabling a photon shot noise limited measurement as demonstrated in figure 3b.

Since the amplitude of the Hahn signal decays with $\tau$ due to decoherence, $\delta S$ likewise decays with $\tau$. $\tau$ itself is limited by the dephasing time $T_2$. Consequently, an optimal $\tau$ exists that maximizes sensitivity and these were determined for the Hahn echo (AC electric field) and the Free Induction Decay (FID) (DC electric field) measurements. The optimal $\tau$ for both cases are included with the plot of $\delta E_{min}$ as a function of $\tau$ in figure 3b. For DC electric fields a maximum $E_{sen}$ of 631.1±15.1V/cm$\sqrt{Hz}$ and a minimum $\delta E_{min}$ of 172.8 V/cm was measured. For AC electric fields a maximum $E_{sen}$ of 142.6 ±3.8V/cm$\sqrt{Hz}$ and a minimum $\delta E_{min}$ of 7.5 V/cm was measured. Since the measured $T_2$ (304 ±36 µs) of the centre investigated was rather short for a NV centre in bulk diamond, an improvement in the measured sensitivity is expected if a NV centre in a higher purity sample was used.

The photon shot-noise limited AC electric field sensitivity of 142.6±3.8V/cm$\sqrt{Hz}$ allows for sensing of an electrostatic field produced by a single elementary charge placed at a distance of 150 nm from the sensing NV spin with one second averaging. An electric field strength as small as 14 V/cm was measured upon averaging for about 100 seconds. This is roughly equivalent to sensing a single electron charge from a distance of 35 nm with a signal to noise ratio of more than 1000. Tailoring the material aspects of diamond,[13] efficient photon collection [27] and multipulse control schemes [28] are options available for further improvement. Importantly, the observed $T_2$ and $T_2^*$ were found to be dependent on the applied axial magnetic field, providing a switch between magnetic and electric field dominated noise regimes and of relevance to decoherence based detection schemes [29]. As shown in Fig. 3d $T_2^*$ strongly depends on the applied axial magnetic field component with a maximum when the applied magnetic field is zero. This is best understood by analysing the eigenstates of Hamiltonian (1) for B, Π =0. The eigenstates of the zero-field Hamiltonian, expressed in terms of the eigenstates of $S_z\{|0\rangle, |+1\rangle, |-1\rangle\}$, are

$$|S_0\rangle = |0\rangle, |S_+\rangle = \frac{1}{\sqrt{2}}\left(e^{-i\phi_\sigma/2}|+1\rangle - e^{i\phi_\sigma/2}|-1\rangle\right), |S_-\rangle = \frac{i}{\sqrt{2}}\left(e^{-\frac{i\phi_\sigma}{2}}\left|+1\right\rangle + e^{\frac{i\phi_\sigma}{2}}\left|-1\right\rangle\right).$$

The expectation value of the spin operator $\langle\vec{S}\rangle$ is zero for each of these eigenstates, which implies that as long as $\mu_B g_e B_z \ll d_{gs}^\perp \sigma_\perp$, there will be no first order Zeeman shift of the energies due to a magnetic field. However, the expectation value of the electric dipole operator $\langle\vec{d}_{gs}\rangle$ is maximal. Therefore, in this regime the decoherence of the ground state spin will be dominated by electric and strain noise and will be minimally effected by magnetic noise. The opposite is true if $\mu_B g_e B_z \gg d_{gs}^\perp \sigma_\perp$, in which case the eigenstates and energies become approximately $\{|0\rangle, |+1\rangle, |-1\rangle\}$ and $E_0 = -(2/3)D_{gs}$ and $E_\pm = (1/3)D_{gs} \pm \mu_B g_e B_z$ respectively. The application of an axial magnetic field thereby switches the decoherence of the ground state spin from being dominated by electric and strain noise to magnetic noise. For the intermediate case $\mu_B g_e B_z \sim d_{gs}^\perp \sigma_\perp$, it is expected that the decoherence will be influenced by each of the source of noise.

Benchmarking the sensitivities of a single NV electric field sensor against other precision charge measuring techniques such as SET, SCM, EFM and Kelvin Probe microscopy shows that the present sensitivity is approximately two orders of magnitude less than the most sensitive electric field detection technique, being the SET at a standoff distance of 100nm. However, the key advantage of using the NV centre comes from the fact that this is anatomic scale sensor, and could perform precision sensing at distances as close as a few nm under ambient conditions. This is in contrast to all of the above mentioned techniques either in terms of spatial resolution or operating conditions. This could readily find application in imaging individual charges with nanometre spatial resolution. In addition the NV sensor has a unique ability to be switched between electric or magnetic field detection modes, turning it into a universal detector system with unprecedented versatility.

**Methods**

In order to accurately align the magnetic field into the non-axial plane ($B_z=0$), the orientation dependent Zeeman splittings of the NV centre's ground state hyperfine structure were observed using CWODMR techniques. The hyperfine interaction of the NV centre's electronic spin with the $^{14}$N nuclear spin results in three observable magnetic transition frequencies separated by 2.2 MHz. Each magnetic transition corresponds to a change in the electronic spin projection $m_s = 0 \leftrightarrow m_s = \pm 1$, but conserve nuclear spin projection, such that the low and high frequency transitions are between hyperfine states with $m_I = \pm 1$ and the central frequency transition is between hyperfine states with $m_I = 0$. For magnetic fields applied to NV centres with crystal strain that satisfies $d_{gs}\sigma/\mu_B g_e B \ll 1$, the hyperfine states with $m_I = \pm 1$ are not split by the presence of strain and/or a non-axial magnetic field, but are split by the axial component of the magnetic field. Consequently, the magnetic field can be aligned in the non-axial plane by manipulating the field and observing the pulsed CW ODMR spectra such that there is no splitting in either the low or high frequency hyperfine magnetic resonance lines. This alignment technique is limited by the line width of these hyperfine lines, but the axial magnetic field component can be determined to a certain uncertainty by interpolating the observed magnetic fields in relation to the coil currents. Once aligned, the magnitude of the non-axial strain $\sigma_\perp$ can be inferred from the splitting of the central line as long as the non-axial magnetic field strength is known, and for the NV centre studied it was found to be $\sigma_\perp = 0.189$ MHz. Given that only the central hyperfine transition which involves states with $m_I = 0$ is susceptible to splitting by strain and electric fields, it was this transition that was used in the electric field measurements reducing the observable ODMR contrast to a third.

Given that the ground triplet state is an orbital singlet and the Stark effect is purely an orbital interaction, the presence of the effect in the ground triplet state is forbidden without some prior mixing of the orbital and spin components of the centre's electronic states. The effect is however allowed without such mixing in the excited triplet state, as it is an orbital doublet. Indeed, the strong linear splitting of the orbitally degenerate $m_s = 0$ spin sub-levels of the excited triplet has been well documented and attributed to the presence of a permanent electric dipole moment $\vec{d}_{es}$. [23] The application of the well established molecular model [24] yields, $\vec{d}_{es} = <e_x|\vec{x}|e_x> + <e_x|\vec{y}|e_y> + <a_1|\vec{z}|a_1> + 3<e_x|\vec{z}|e_x>$ where $a_1$, $e_x$, and $e_y$ are the molecular orbitals that transform as the $A_1$ and $E$ irreducible representations of the centre's $C_{3v}$ symmetry group constructed from the dangling $sp^3$ atomic orbitals of the nearest neighbours of the vacancy. The charge distributions $e_x e_x$ and $e_x e_y$ responsible for the dominant non-axial components of $\vec{d}_{es}$ are depicted in supplementary figure 1. The previous observation of the weak linear splitting of the degenerate $m_s = \pm 1$ spin sub-levels of the ground triplet state [24] suggests that the Stark effect in the ground triplet state is indeed a consequence of a higher-order interaction. Spin-orbit interaction is the immediately obvious higher-order interaction capable of the required mixing, and the application of the molecular model indicates that the $m_s = \pm 1$ spin sub-levels of the ground triplet state are mixed with the degenerate $m_s = 0$ spin sub-levels of the excited triplet state by the non-axial components of spin-orbit interaction. This mixing between the ground and excited triplet states explains the existence of the weaker linear Stark effect in the ground triplet state by the ground triplet states' permanent electric dipole moment $\vec{d}_{gs}$ being related to the dipole moment of the excited triplet state by the square of the non-axial spin-orbit interaction strength $\lambda_\perp$, such that $\vec{d}_{gs} \propto (\lambda_\perp^2/E_{es}^2)\vec{d}_{es}$, where $E_{es}$ is the energy of the excited triplet state.

**Acknowledgments**

This work was supported by the EU (QAP, EQUIND, NEDQIT, SOLID), DFG(SFB/TR21, FOR730 and FOR1482), NIH, Landesstiftung BW, BMBF (EPHQUAM,KEPHOSI), Volkswagen Stiftung, and the Australian Research Council TN wishes to thank Vienna doctoral program CoQuS(Austrian Science Fund (FWF) project W1210).

**Author Contributions**

F.D, H.F,T.N, G.B, T.W, F.J. carried out the experiments; M.W.D., F.R., L.C.L.H developed the theory. All authors discussed the results, analyzed the data and commented on the manuscript.
J.W. wrote the paper and supervised the project.

**Competing Financial Interests**

The authors declare no competing financial interests.

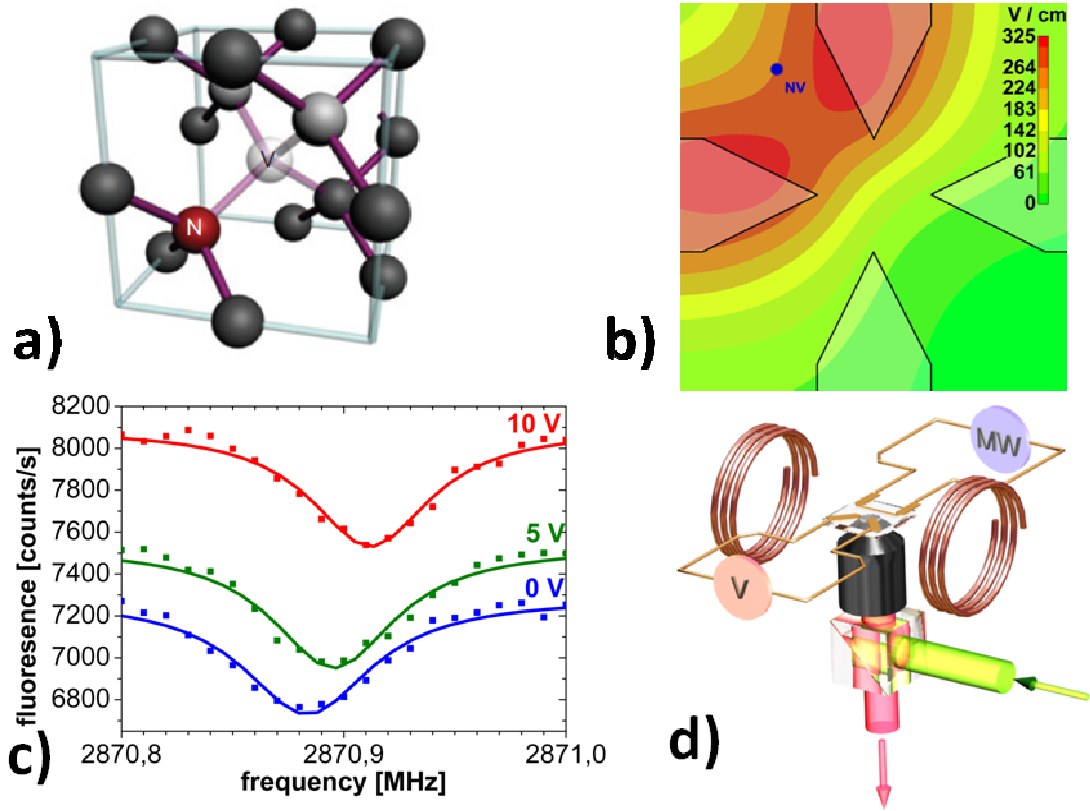

**Figure 1 Schematic of the NV and the measurement scheme a,** Schematic drawing of the NV centre with one nitrogen at a carbon lattice site and an adjacent vacancy **b,** simulated absolute electric field 6 µm below the microstructure (depth of the NV) for 1 V voltage difference applied **c,** observed shift of the ODMR resonance lines for different voltages applied to the electrodes. One can clearly see the effect of a Stark shift **d,** schematic of the used confocal setup with Helmholtz coils for magnetic field alignment and a microstructure on the diamond sample to create the electric field and couple in the microwave

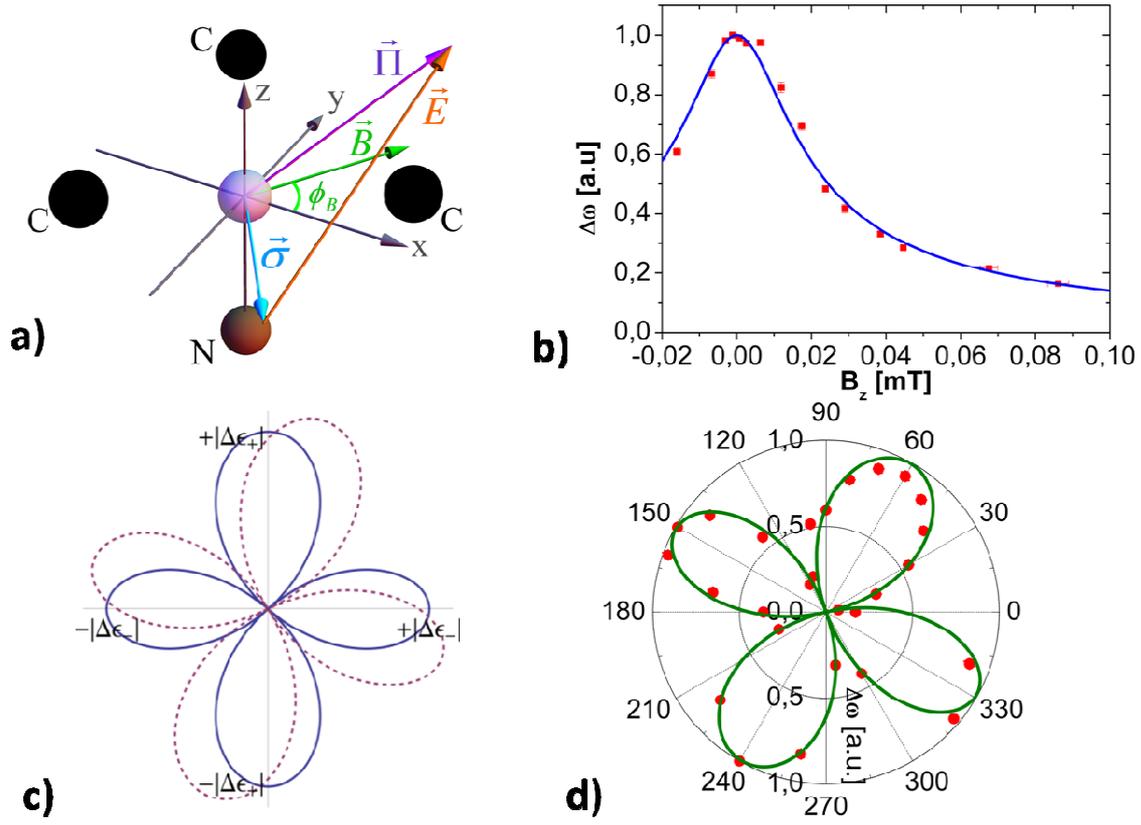

**Figure 2 Theory of the NV electric field sensing and measured results. a,** schematics of the NV centre, coordinate axes and the magnetic, electric and strain fields defined in the text. The solid spheres represent the nuclei of the respective atoms neighbouring the vacancy (transparent). The coordinate axes are defined such that the z axis coincides with the axis of symmetry connecting the nitrogen and vacancy sites. **b,** The measured normalised magnetic transition frequency change **Δω** due to an applied AC electric field as a function of the axial magnetic field strength. The blue solid line is the theoretical fit using equation (2). **c,** Theoretical change in the magnetic transition frequency **Δω** due to an applied electric field as a polar function of the magnetic field orientation $\phi_B$ in the non-axial plane. The blue line corresponds to the case where the applied electric field and the effective strain field are parallel, and the red dashed line corresponds to a **10°** rotation of the external electric field with respect to the strain field. The extremities of the parallel case are defined by $\Delta\varepsilon_\pm = d_{gs}^\parallel E_z \pm d_{gs}^\perp E_\perp$. **d,** Polar plot of the measured **Δω** as a function of the magnetic field orientation $\phi_B$ in the non-axial plane. The green solid line is the theoretical fit using equation (2).

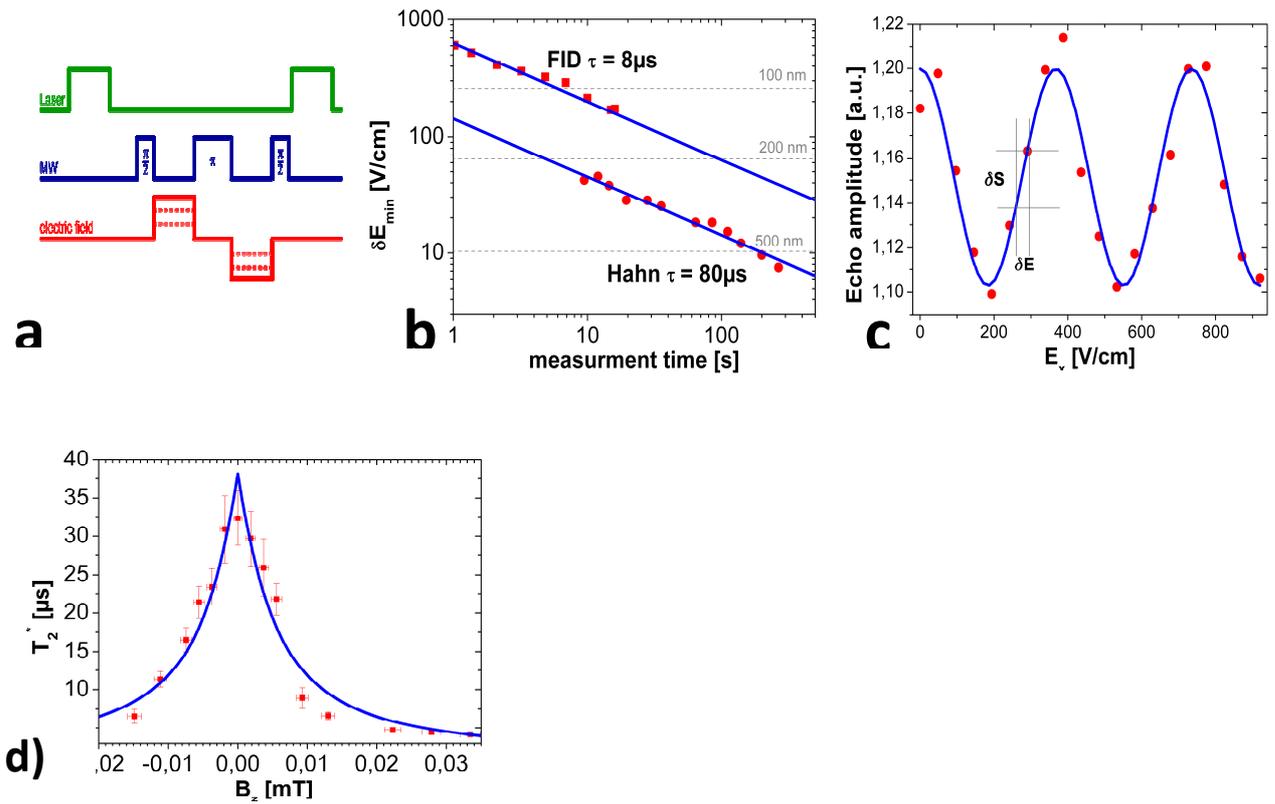

Figure 3 Sensitivity and coherence time measurements. a, Hahn echo pulse scheme with alternating square electric field for the AC measurements (for the DC sequence see supplementary information) b, The measured minimal detectable change in the electric field $\delta E_{min}$ as a function of the measurement time. The solid line corresponds to the shot noise limit. c, Measured optical Hahn signal as a function of the applied non-axial electric field strength. The solid line is a fit using cosine function (see supplementary information) and a fixed free evolution time, $\tau = (8), 80$ µs, in the (FID), Hahn echo sequence. d) Measured dependence of the NV centre's coherence time $T_2^\star$ on the axial magnetic field. The solid line shows a fit of the data (see supplementary information).

**Supplementary Information**

**The Electric Dipole Moments of the Ground State Triplet**

As discussed in methods, the application of the molecular model of the centre [24] yields the electric dipole moment of the ground and excited triplet states in terms of the molecular orbitals of the centre

$$\vec{d}_{es} \propto \frac{\lambda_\perp^2}{E_{es}} \vec{d}_{es} = <e_x|\vec{x}|e_x> + <e_x|\vec{y}|e_y> + <a_1|\vec{z}|a_1> + 3<e_x|\vec{z}|e_x>$$

where the molecular orbitals are defined in terms of the tetrahedrally coordinated dangling $sp^3$ atomic orbitals of the nearest neighbour carbon atoms of the vacancy $(c_1, c_2, c_3)$ to be

$$a_1 = \frac{1}{\sqrt{3}\sqrt{1+2S_{cc}}}(c_1 + c_2 + c_3)$$

$$e_x = \frac{1}{\sqrt{3}\sqrt{2-2S_{cc}}}(2c_1 - c_2 - c_3)$$

$$e_y = \frac{1}{\sqrt{2-2S_{cc}}}(c_2 - c_3)$$

with $S_{cc} = <c_1|c_2>$ being the overlap integral of the atomic orbitals. Although the molecular orbitals are approximations of the true orbitals, they provide a simple model that is qualitatively correct. Supplementary figure 1 depicts plots of the charge distributions $e_x e_x$ and $e_x e_y$ responsible for the dominant non-axial components of $\vec{d}_{es}$ and $\vec{d}_{gs}$. The plots provide a clear visualisation of the charge separations that yield the x and y components of the dipole moments.

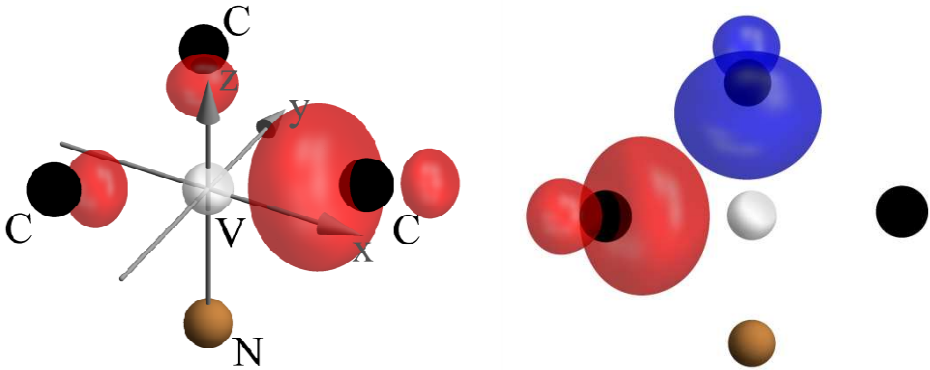

**Supplementary figure 1.** Electronic charge distributions responsible for the x (left) and y (right) components of the electric dipole moments of the ground and excited triplet states. The solid spheres represent the nuclei of the respective atoms neighbouring the vacancy (transparent). The positive and negative contributions to the charge distributions are red and blue respectively. The coordinate axes are defined such that the z axis coincides with the axis of symmetry connecting the nitrogen and vacancy sites.

**Pulsed Optically Detected Magnetic Resonance Techniques**

The electric field sensing measurements were conducted using optically detected magnetic resonance (ODMR) techniques in conjunction with applied electric and magnetic fields. The applied electric field was generated by applying a controlled voltage to electrodes within a microstructure fabricated by lithography and electroplating directly on a bulk sample of diamond containing NV centres. A particular NV centre between the electrodes was selected and used for all of the measurements The resonant microwave field used in conjunction with the optical field in implementing the ODMR techniques was generated by the gold coils adjacent to the electrodes in the microstructure with the signal carried by the outer wire and the ground plane in the inner wire. Two Helmholtz coil pairs for the x-y axes and a single coil for the z axis were used to generate and precisely manipulate the applied magnetic field (as described in methods and below). A 532 nm wavelength laser and a standard confocal arrangement were used to perform the optical spin polarization and readout of the ground state spin as part of the ODMR techniques.

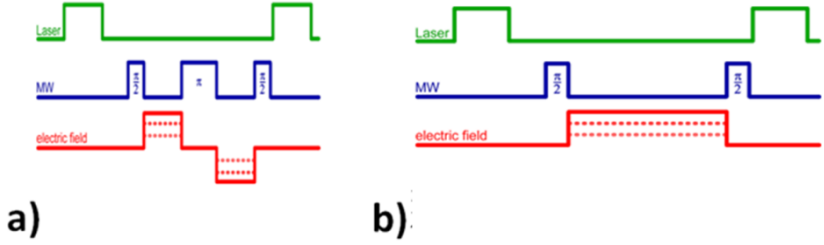

**Supplementary figure 2 ODMR pulse sequences. a, Hahn** echo pulse sequence used for AC electric field measurements. **b,** FID pulse sequences used for the DC electric field measurements.

The AC and DC electric field measurements were conducted using Hahn echo and Free Induction Decay (FID) ODMR pulse sequences (refer to supplementary figure 2). Each sequence used an optical pulse to initialize and readout the ground state spin. The microwave pulses performed the $\pi$ and $\pi/2$ rotations of the spin as part of the magnetic resonance sequences and AC and DC square voltage signals were used to generate square electric field signals between the microwave pulses. Note that the final microwave pulses of each sequence project the spin state into a population difference which can be detected by means of a fluorescence intensity difference (as part of the optical readout). The principle parameters of the sequences were the free evolution time between pulses $\tau$ and the voltage signal amplitude which governs the applied electric field strength.

The collected phase of the spin during the FID and Hahn echo sequences are respectively

$\Phi_{\text{FID}} = \int_0^\tau \Delta\omega(t)\, dt$  (1)
and

$\Phi_{\text{HE}} = \int_0^\tau \Delta\omega(t)\, dt - \int_\tau^{2\tau} \Delta\omega(t)\, dt$  (2)

which result in different oscillating signals given by

$\Delta I(\Delta\omega, \tau) = A \cos\Phi$  (3)

With $\Delta\omega = 2\pi d_{gs}^\perp E_\perp$, a rectangular electric field pulse shape and no phase difference between the electric field and the pulse sequence, it follows that

$\Delta I_{\text{FID}}(E_\perp, \tau) = A \cos(2\pi d_{gs}^\perp E_\perp \tau)$  (4)

and

$\Delta I_{\text{HE}}(E_\perp, \tau) = A \cos(4\pi d_{gs}^\perp E_\perp \tau)$  (5)

where $\Delta I$ is the amplitude change of the optically detected signal, A is the ODMR contrast, $E_\perp$ is the non-axial electric field strength and $d_{gs}^\perp = 17 \pm 2.5$ Hz cm/V [23]. For a fixed $\tau$ and an increasing $E_\perp$ an oscillation in the detected fluorescence signal can be detected. In order to collect the maximum phase difference the alternating electric field has to be phase matched with τ so that the π pulse corresponds with the inversion of the alternating field. In order to detect an alternating field with a random phase, one has to conduct a series of measurements with the same measurement parameters, but a defined phase difference between the single measurements. From the measured phase dependent oscillations, one can determine the electric field strength. In this work only alternating fields with a perfect phase match were investigated.

**Alignment of the Applied Magnetic Field**

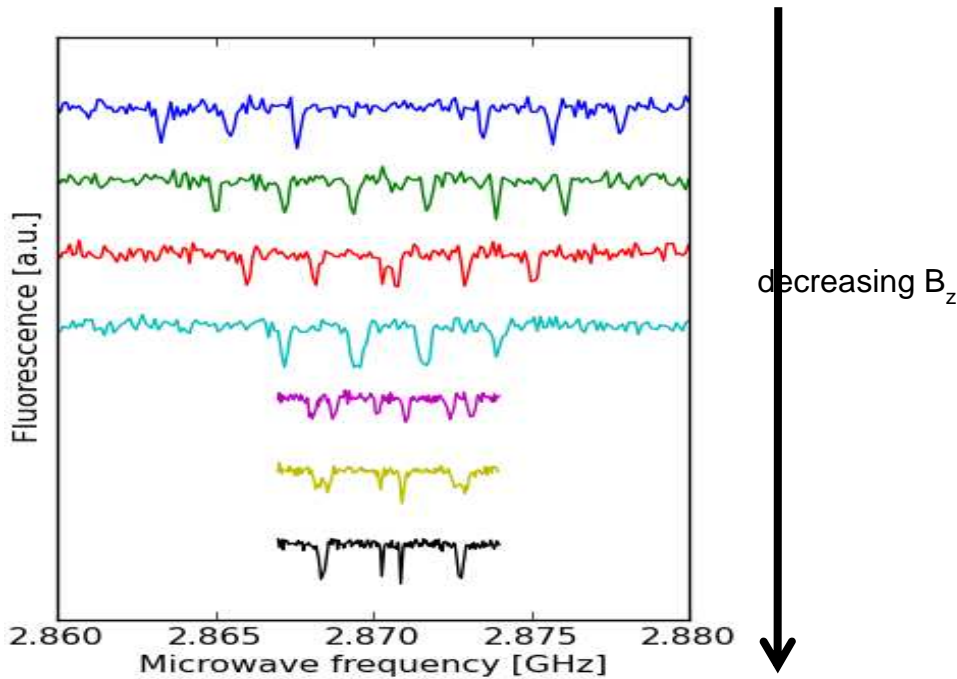

**Supplementary figure 3** Continuous wave ODMR spectra used to conduct magnetic field alignment. With decreasing $B_z$ only the central lines ($m_i=0$) are split by $\sigma_\perp$. The distance between the outer lines were used to calculate the remaining $B_z$.

As discussed in methods, the alignment of the applied magnetic field was conducted by observing the hyperfine spectra of the centre using continuous wave ODMR techniques. Due to the small strain present in the used NV centre, the hyperfine interaction prevents a linear Stark shift of the transitions associated with the $m_i=\pm 1$ states for zero $B_z$, thereby allowing us to use these transitions for magnetic field alignment.

For a small non-axial magnetic field $B_\perp$, the axial magnetic field is given by $B_z = \frac{\Delta\vartheta}{g\mu_B}$, where $\Delta\vartheta$ is the observed frequency difference between the $m_s=0$ to $m_s=+1$ and $m_s=0$ to $m_s=-1$ transitions. Since the hyperfine interaction creates a threefold of these transitions one has to consider the frequency difference between the lines that observe the same frequency shift due to the hyperfine interaction ( e.g. $|m_s=+1\rangle|m_i=+1\rangle$ and $|m_s=-1\rangle|m_i=-1\rangle$). For zero $B_z$ these transitions should overlap perfectly. The minimal detectable magnetic field is here determined by the line width of the observed transition. In order to align the magnetic field, we sweep the current applied to the coils, thereby sweeping $B_z$. In supplementary figure 3 a schematic of this sweep is shown. From interpolating the magnetic field in relation to the current in the coil, we were able to extrapolate the magnetic field values where the observed lines overlap.

For larger $B_\perp$ a simple determination of $B_z$ is not possible any more, since the $m_s=\pm 1$ states are mixed by $B_\perp$. This leads to a decrease of the observed hyperfine splitting and a decrease of the observed line shift due to $B_z$. Fortunately the lines still show an overlap for zero $B_z$, which was used to align the magnetic field in the high $B_\perp$ regime. This leads to an alignment error of approx. 0.003 mT and therefore to an error in the interaction strength of about 10 percent (see figure 2b).

**Reduction of $T_2^*$ at small magnetic fields**

As depicted in the main text (Fig. 3d), a dependence of the NV centre's coherence time $T_2^*$ on the axial component of the magnetic field was observed. The $T_2^*$ clearly obtained a maximum value when the applied axial magnetic field was reduced to zero.

Given that the magnetic noise present in the diamond crystal, produced by such sources as the $^{13}$C spin bath, is much greater than the electric noise, the application of an axial magnetic field reduces the $T_2^*$ of the centre's ground state spin. The dependence of $T_2^*(B_z)$ on the axial magnetic field was modelled by the simple expression

$$T_2^*(B_z) = \kappa(B_z)\left(T_2^{*,\perp} - T_2^{*,\parallel}\right) + T_2^{*,\parallel} \quad (6)$$

where $\kappa(B_z)$ is the mixing of $|+1\rangle$ and $|-1\rangle$ as obtained from the eigenstate solution of the spin-Hamiltonian in the presence of an applied magnetic field, and $T_2^{*,\perp}$ and $T_2^{*,\parallel}$ are the coherence times for a zero axial magnetic field and a given maximum axial magnetic field $B_{z,max}$ respectively, such that $T_2^{*,\parallel} \ll T_2^{*,\perp}$. This model was used to fit the observed dependence of $T_2^*$ as depicted in supplementary figure 3. Regardless of its simple nature, the model has produced a reasonable fit of the data. Hence, it appears that a precisely controlled magnetic field can switch the centre between regimes in which either the magnetic or electric field noise is dominant. These observations have potentially important implications for the application of the NV centre in QIP (e.g. switching the interaction between two close NVs) and decoherence imaging.